\documentclass[amsfonts,amsmath,amssymb,showpacs,floatfix,superscriptaddress,nofootinbib,preprintnumbers,11pt]{article}
\usepackage{amsmath,amsthm,amssymb}
\usepackage{graphicx}
\begin{document}
\title{\textbf{\textsf{Interacting New Generalized Chaplygin Gas}}}
\author{ \textsf{Mubasher Jamil\footnote{mjamil@camp.edu.pk}}
 \\ \\
\small Center for Advanced Mathematics and Physics, Campus of
College of E\&ME,\\ \small National University of Sciences and
Technology, \small Peshawar Road, Rawalpindi - 46000, Pakistan
 }\maketitle
\begin{abstract}
We have presented a model in which the new generalized Chaplygin gas
interacts with matter. We find that there exists a stable scaling
solution at late times in the evolution of the universe. Moreover,
the phantom crossing scenario is observed in this model.
\end{abstract}
 \large
\newpage
\section{Introduction}
Astrophysical observations show that more then seventy percent of
the cosmic energy density is contained in an unknown `dark' sector
commonly termed as `dark energy' \cite{perl} (see \cite{pad} for
reviews on dark energy). The remaining part of the total energy
density is due to matter, which is also mostly dark \cite{bachall}.
This mysterious dark energy is specified by an equation of state
(EoS) $p_{de}=\omega_X\rho_{de}$, where $p_{de}$ and $\rho_{de}$ are
the pressure and energy density of dark energy, while $\omega_X$ is
the corresponding dimensionless EoS parameter. In the presence of
dark energy, the fabric of spacetime expands in an accelerated
manner, implying $p_{de}<0$ and consequently $\omega_X<0$. Since
general relativity cannot satisfactorily explain the cosmic
accelerated expansion, it motivates theorists either modifying
curvature or the matter part in the Einstein field equations. In
recent years, such options are carefully investigated in literature.
Alternative gravity theories based on the modification of
Einstein-Hilbert action include $f(R)$ gravity \cite{nojiri},
scalar-tensor gravity \cite{jarv} and Lovelock gravity \cite{love},
to name a few. Similarly there are models in which only the matter
term is modified, examples are Cardassian model \cite{card}, the
bulk viscous stress \cite{brevik} and the anisotropic stress
\cite{wands}. However all such modified models have several
drawbacks in explaining the observational data. There is also a
possibility that the dark energy admits an exotic EoS that manifests
the observed accelerated expansion. We consider such a possibility
by introducing an EoS based on the Chaplygin gas (CG). In the
context of cosmology, the Chaplygin gas was first introduced by
Kamenshchik et al \cite{kamen}. It is specified by $p=-L/\rho$,
where $L$ is a constant. Its density evolution is given by
\begin{equation}
\rho=\sqrt{L+M (1+z)^6},
\end{equation}
where $M$ is constant of integration and $z$ is the redshift
parameter. Interest in CG arose when it appeared that it gives a
unified picture of dark energy and dark matter i.e. under certain
constraints on $L$ and $M$, expression (1) gives density evolution
of matter at high redshifts and dark energy at low redshifts
\cite{bilic}. Other successes of CG is that it explains the recent
phantom divide crossing \cite{zhang}, is consistent with the data of
type Ia supernova \cite{senn} and the cosmic microwave background
\cite{liu}. The CG emerges as an effective fluid associated with
$d$-branes \cite{borde} and can also be obtained from the
Born-Infeld action \cite{bento}. Since matter and dark energy are
the dominant components of the cosmic composition, it is natural to
expect their mutual interaction at some scale. The exact nature of
this interaction is still unexplained and the interaction may not
necessarily be gravitational either. Cosmological models based on
the interaction between dark energy and matter are termed
`interacting dark energy' in literature and are under thorough
investigation \cite{lima}. In these models, either cosmic specie
decays into the other depending on the sign of the coupling
parameter involved. Recent interest in interacting dark energy
models is also triggered from the astrophysical observations which
show that the energy densities of matter and dark energy are of the
same order of magnitude i.e. $r_m\equiv\rho_m/\rho_{de}\simeq1$. It
leads to the `cosmic coincidence problem' which asks the explanation
of $r_m\simeq1$ at the present time. Alternatively, why the dark
energy parameter $\omega_X$ is close to $-1$ in recent times. In the interacting dark energy scenario, there has been successful attempts
in resolving this problem and stable attractor solutions of the
Friedmann-Robertson-Walker (FRW) equations are obtained which give
$\omega_X\simeq-1$ closer to present time \cite{quartin}. It is shown in \cite{marek} that the cubic corrections to the Hubble law, measured by distant supernovae type Ia, probes this interaction. Moreover, this interaction is controlled by third and higher derivatives of the scale factor. Moreover, cosmic microwave background observations lead to a constraint on the coupling parameter of the interaction as $0.05<c<0.2$ \cite{wang22}. Observationally the Abell cluster A586 provides evidence of the interaction between dark matter and dark energy \cite{orfeu}.

In the context of field theory and particle physics, it is customary and appealing to interpret the dark energy as some sort of particles that interact with the particles of the standard model very weakly. The weakness of the interaction is required since dark energy particles have not been produced in the accelerators and because dark energy has not yet been decayed into lighter or massless fields such as photons. The interaction between dark energy and other particles cannot be arbitrary since this interaction gives a fifth force with a range $\lambda\sim1/m_0$, where $m_0$ is the the mass of dark energy particle. It has been shown in \cite{mac} that an equation of state with $\omega_{de}<-1$ can be a signal that dark energy will decay in the future and the universe will stop accelerating. This conclusion is based in interpreting a $\omega_{de}<-1$ as a signal of dark energy interaction with another fluid. In another paper \cite{sean}, it is shown that the mass of dark energy particle could be of the order $m_0^2\sim(10^{-33}\text{eV})^2$. In the same study, it is proposed that the phantom particle can decay into one or more phantom plus an ordinary baryonic particle. Moreover, an ordinary particle may decay into phantoms plus other ordinary particles with a larger effective mass than the original. Thus the above discussion shows that model of interacting dark energy is supported by both theoretical arguments and observational evidences.

Zhang and Zhu \cite{zhang22} used the Chaplygin gas in the interacting dark
energy model and obtained the stable scaling solution of the FRW
equations. Later on, their work was extended by Wu and Yu \cite{wu}
for the generalized Chaplygin gas. We here extend these earlier
studies by using the new generalized Chaplygin gas.
\section{Modeling of dynamical system}
We start by assuming the background to be spatially homogeneous and
isotropic FRW spacetime
\begin{equation}
ds^2=-dt^2+a^2(t)[dr^2+ r^2(d\theta^2+\sin^2\theta d\phi^2)].
\end{equation}
The equations of motion corresponding to FRW spacetime filled with
the two component fluid are
\begin{eqnarray}
\dot{H}&=&-\frac{\kappa ^{2}}{6}(p_{ncg}+\rho _{ncg}+\rho _{m}),\\
H^{2}&=&\frac{\kappa ^{2}}{3}(\rho _{ncg}+\rho _{m}).
\end{eqnarray}
Here $\kappa ^{2}=8\pi G$ is the Einstein's gravitational constant
and $H=H(t)\equiv\dot a/a$ is the Hubble parameter. In this paper,
we solve the FRW equations using the `new generalized Chaplygin gas'
(NCG) state equation, proposed by Zhang et al \cite{zhang1}. It is
an extended form of the generalized Chaplygin gas and hence dubbed
with the `new'. The NCG model is dual to an interacting XCDM
parameterization scenario, in which the interaction is determined
by the parameter $\alpha$. Here the X part corresponds to the
quintessence ($\omega_X<-1/3$), following the notation used in
\cite{zhang1}. Since the observational data favors $\omega_X$ to be
in the range ($-1.46,-0.78$) \cite{perl,pad}, it has motivated to
generalize the Chaplygin gas EoS to the NCG form to incorporate any
X-type dark energy in the universe.
\begin{equation}
p_{ncg}=-\frac{\tilde{A}(a)}{\rho _{ncg}^{\alpha }},\ \
\tilde{A}(a)=-\omega_XAa^{-3(1+\omega_X)(1+\alpha)}.
\end{equation}
The density evolution of NCG is given by
\begin{equation}
\rho
_{ncg}=[Aa^{-3(1+\omega_X)(1+\alpha)}+Ba^{-3(1+\alpha)}]^{\frac{1}{1+\alpha}},
\end{equation}
where $B$ is the constant of integration. The energy conservation
equation for the dynamical system under consideration is
\begin{equation}
\dot\rho_{ncg}+\dot\rho_m+3H(\rho_m+\rho_{ncg}+p_{ncg})=0.
\end{equation}
Due to interaction between the two components, the energy
conservation would not hold for the individual components, therefore
the above conservation equation will break into two non-conserving
equations:
\begin{eqnarray}
\dot{\rho }_{ncg}+3H(p_{ncg}+\rho _{ncg})&=&-Q,\\
\dot{\rho }_{m}+3H\rho _{m}&=&Q.
\end{eqnarray}
Here $Q$ is the energy exchange term which is to be specified ad
hoc. However from the dimensional considerations, it is obvious that
$Q$ should have dimensions of density into inverse time, the later
being chosen to be Hubble parameter. Thus we expect that
$Q=Q(H\rho_{ncg},H\rho_m)$ which upon expanding about densities in a
Taylor series yields $Q\simeq H(\rho_{ncg}+\rho_m)$ \cite{setare11}.
We also insert a coupling parameter $c$ in $Q$ to determine the
strength of the interaction, thus we have
\begin{equation}
Q=3Hc(\rho _{ncg}+\rho _{m}).
\end{equation}
From the observational data of 182 Gold type Ia supernova samples,
CMB data from the three year WMAP survey and the baryonic acoustic
oscillations from the Sloan Digital Sky Survey, it is estimated that
the coupling parameter between dark matter and dark energy must be a
small positive value (of the order unity), which satisfies the
requirement for solving the cosmic coincidence problem and the
second law of thermodynamics \cite{feng}. The positive $c$ implies
that the energy will flow from the NCG into matter. To study the
dynamics of our system, we proceed by setting
\begin{equation}
x=\ln a=-\ln (1+z),
\end{equation}
which is termed as the e-folding time parameter and $z$ is the redshift
parameter. Moreover, the density and pressure of NCG can be
expressed by dimensionless variables $u$ and $v$ as
\begin{equation}
u=\Omega _{ncg}=\frac{\rho _{ncg}}{\rho _{cr}}=\frac{\kappa ^{2}\rho
_{ncg}}{ 3H^{2}},\ \ v=\frac{\kappa ^{2}p_{ncg}}{3H^{2}}.
\end{equation}
The EoS parameter $\omega_X $ is conventionally defined by
\begin{equation}
\omega_X (x)\equiv \frac{p_{ncg}}{\rho _{ncg}},
\end{equation}
which after using (12), becomes
\begin{equation}
\omega_X(x)=\frac{v}{u}.
\end{equation}
The density parameters of NCG and dark matter are related as
\begin{equation}
\Omega _{m}=\frac{\kappa ^{2}\rho _{m}}{3H^{2}}=1-\Omega _{ncg}=1-u.
\end{equation}
Using Eqs. (12),(13) and (15) in (8) and (9), we obtain
\begin{eqnarray}
\frac{du}{dx}&=&-3c-3v+3uv,\\
\frac{dv}{dx}&=&3\frac{\alpha
v}{u}(u+v+c)+3v(1+\alpha)(1+\omega_X)+3v(1+v).
\end{eqnarray}
The critical points of the above system are obtained by equating Eqs. (16) and (17) to zero. The only critical point of the system is
\begin{eqnarray}
u_{c}&=&\frac{2-c+\omega_X}{2+\omega_X},\\
v_{c}&=&-(2+\omega_X).
\end{eqnarray}
Notice that for $\omega_X=-1$, our results reduce to those of
\cite{wu} for the interacting generalized Chaplygin gas. Since in a
spatially flat universe, the meaningful range is $0\leq u\leq1$,
consequently $0\leq u_c\leq1$ implying $0\leq c\leq2+\omega_X$.
Since $\omega_X<0$, therefore $c\geq1$ can arise in certain cases.
Note that the acceleration in the late evolution of the universe
arises when
\begin{equation}
q=-\frac{\ddot{a}}{aH^2}\leq-1,
\end{equation}
Using Eqs. (12) to (15) in (20), one can write
\begin{equation}
q=\frac{1}{2}(1+3v).
\end{equation}
Using (19) in (22), we get
\begin{equation}
q=\frac{1}{2}(1-3(2+\omega_X)).
\end{equation}
For $-1.6<\omega_X<0$, we notice that $q<0$, thus giving accelerated
expansion of the universe (see Fig. 6). We further check the
stability of the dynamical system (Eqs. 16 and 17) about the
critical point ($u_c,v_c$). To do this, we linearize the governing
equations about the critical point i.e. $u=u_c+\delta u$ and
$v=v_c+\delta v$, we obtain
\begin{eqnarray}
\frac{d\delta u}{dx}&=&3v_c\delta u+3(-1+u_c)\delta v,\\
\frac{d\delta v}{dx}&=&\Big[ \frac{-3\alpha}{u_c^2}(v^2_c+cv_c)
\Big]\delta u\nonumber
\\&\;&+\Big[3(1+2v_c+\alpha)+\frac{6\alpha
v_c}{u_c}+\frac{3\alpha c}{u_c}+3(1+\alpha)(1+\omega_X)\Big]\delta
v.
\end{eqnarray}
The eigenvalues corresponding to the linearized system are
\begin{eqnarray}
\lambda&=&\frac{1}{2(-2+c-\omega_X)}[-6c(2+\omega_X)+3(2+\alpha)(2+\omega_X)^2\nonumber
\\&\;&
-3\sqrt{\alpha(2+\omega_X)^2(-4c^2+4c(2+\omega_X)+\alpha(2+\omega_X)^2)}],\\
\mu&=&\frac{1}{2(-2+c-\omega_X)}[-6c(2+\omega_X)+3(2+\alpha)(2+\omega_X)^2\nonumber
\\&\;&
+3\sqrt{\alpha(2+\omega_X)^2(-4c^2+4c(2+\omega_X)+\alpha(2+\omega_X)^2)}].
\end{eqnarray}
To obtain stable critical point, the real parts of the eigenvalues
must be negative. We notice that for $\alpha\geq0$, $0<c<2+\omega_X$
and $\omega_X<0$, the two eigenvalues $\lambda$ and $\mu$ will be
negative, giving a stable attractor solution at ($u_c,v_c$) in the
late time evolution governed by the FRW equations. For instance,
taking $c=0.3$, $\omega_X=-1.2$ and $\alpha=0.2$ (generalized
phantom energy), we have $\lambda=-1.86$ and $\mu=-3.69$; whilst for
$\alpha=1$ (phantom energy), we have $\lambda=-1.64$ and
$\mu=-6.99$. The negativity of the eigenvalues can also be proven as
\begin{equation}
-6c(2+\omega_X)+3(2+\alpha)(2+\omega_X)^2=3(2+\omega_X)(-2c+(2+\alpha)(2+\omega_X))\geq0,
\end{equation}
since $\alpha\geq0$, $0<c<2+\omega_X$ and $\omega_X<0$. Now notice that both the eigenvalues will be
negative if
\begin{eqnarray}
(-6c(2+\omega_X)+3(2+\alpha)(2+\omega_X)^2\geq
9\alpha(2+\omega_X)^2(-4c^2+4c(2+\omega_X)+\alpha(2+\omega_X)^2).
\end{eqnarray}
which yields
\begin{equation}
36(1+\alpha)(2+\omega_X-c)^2\geq0.
\end{equation}
Therefore both the eigenvalues are indeed negative. In Figures 1 and 2, we have provided a pictorial
relationship between $u$ and $v$ for suitable choices of model
parameters and initial conditions. It is interesting to notice that
solutions converge to the same state under the given conditions.
Figures 3 and 4 show that the functions $u$ and $v$ are stable
solutions of the system i.e. the corresponding curves become flat at
approximately $x\simeq3$ and become globally flat afterwards.
Initially these functions are unstable for small values of e-folding
time $x$ but in the late times, these become globally stable. It can
also be seen that NCG state parameter $\omega_X=v/u$ crosses
$\omega_X=-1$ at present time ($x=0$) and enters the phantom regime
$\omega_X<-1$ for large e-folding time (see Fig.5).

\section{Conclusion}
Model of interacting dark energy possesses enormous potential in
resolving or at least simplifying several cosmological problems
including the coincidence problem and the phantom crossing. In this
model, these problems are analyzed by considering two major cosmic
components namely dark matter and dark energy and assuming an
interaction between them. Two non-conserving equations are derived
and then non-dimensionalized. Due to non-linearity in the governing
equations, we sought for numerical solution of the system. The
stability of the solution is determined by performing stability
analysis. In this paper, we have studied an interacting dark energy
model dealing with the interaction between new generalized Chaplygin
gas and the matter. The analysis is performed by doing the stability
check of the dynamical system. We found that the dynamical system is
stable about the only critical point of the system. This shows that
stable stationary attractor solution exists in the late time
evolution of the universe when the later one enters a steady state.
Finally, this paper presents an extension of earlier work in
\cite{wu}.

\subsubsection*{Acknowledgment}
I would like to thank anonymous referees for their useful criticism on this work.

\newpage
\begin{figure}
\includegraphics[scale=.7]{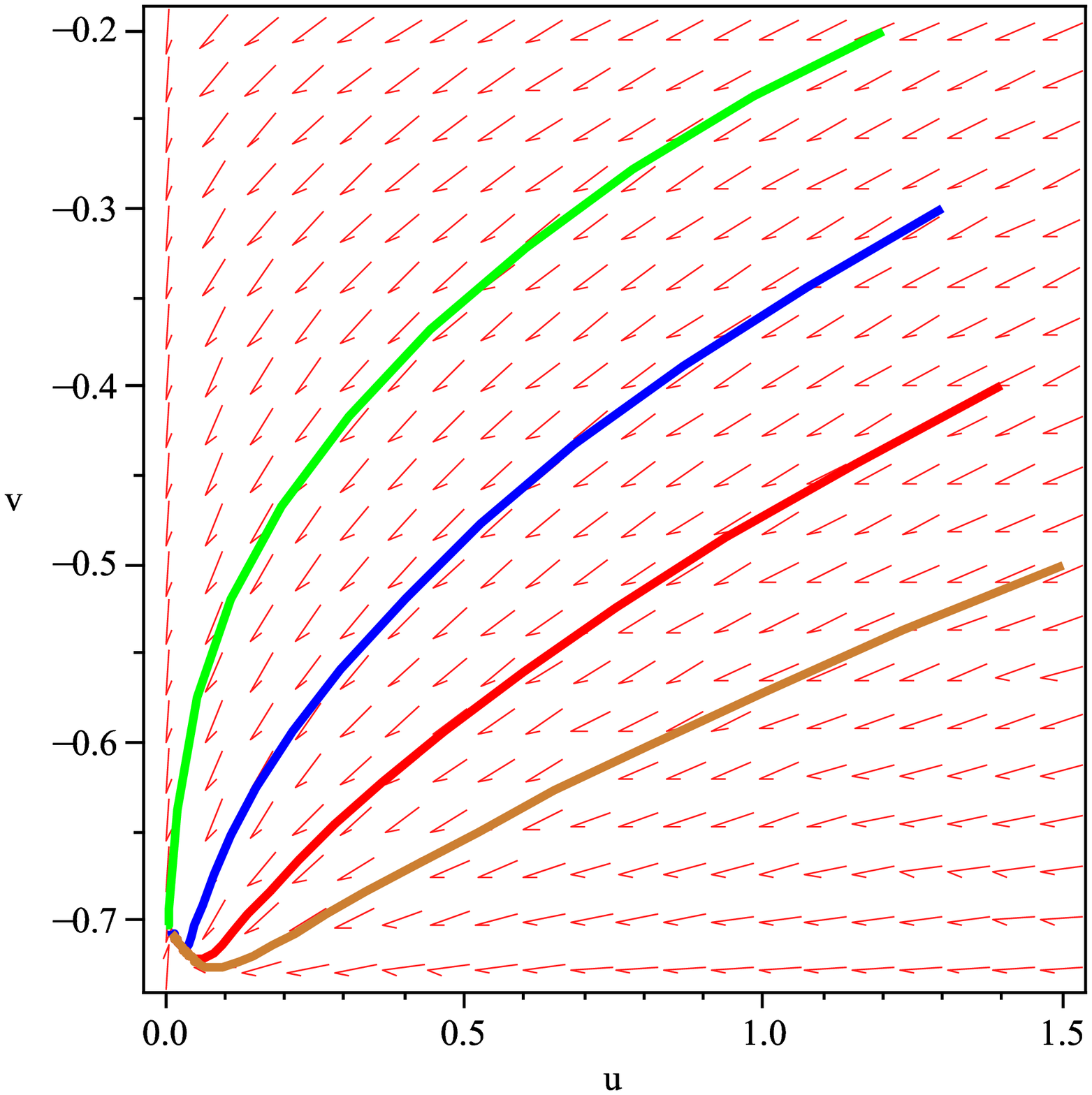}\\
\caption{The phase diagram of the interacting new generalized
Chaplygin gas model with $c=0.7$, $\alpha=0.07$ and $\omega_X=-1.3$.
The curves correspond to the initial conditions $u(-2)=1.2,
v(-2)=-0.2$ (green); $u(-2)=1.3, v(-2)=-0.3$ (blue); $u(-2)=1.4,
v(-2)=-0.4$ (red); $u(-2)=1.5, v(-2)=-0.5$ (brown).}
\end{figure}
\newpage
\begin{figure}
\includegraphics[scale=.7]{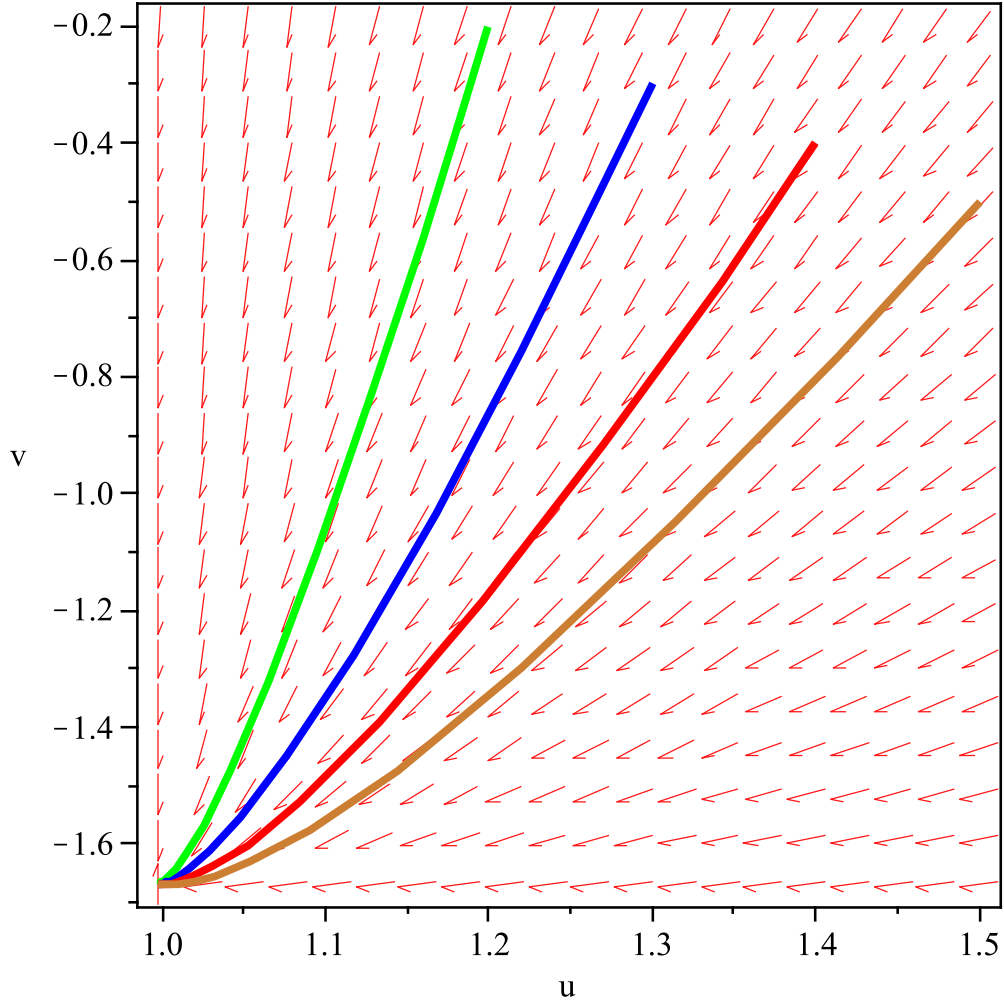}\\
\caption{The phase diagram of the interacting new generalized
Chaplygin gas model with $c=0.003$, $\alpha=0.3$, $\omega_X=-1/3$.
The curves correspond to the initial conditions as given in Fig. 1.}
\end{figure}
\newpage
\begin{figure}
\includegraphics[scale=.7]{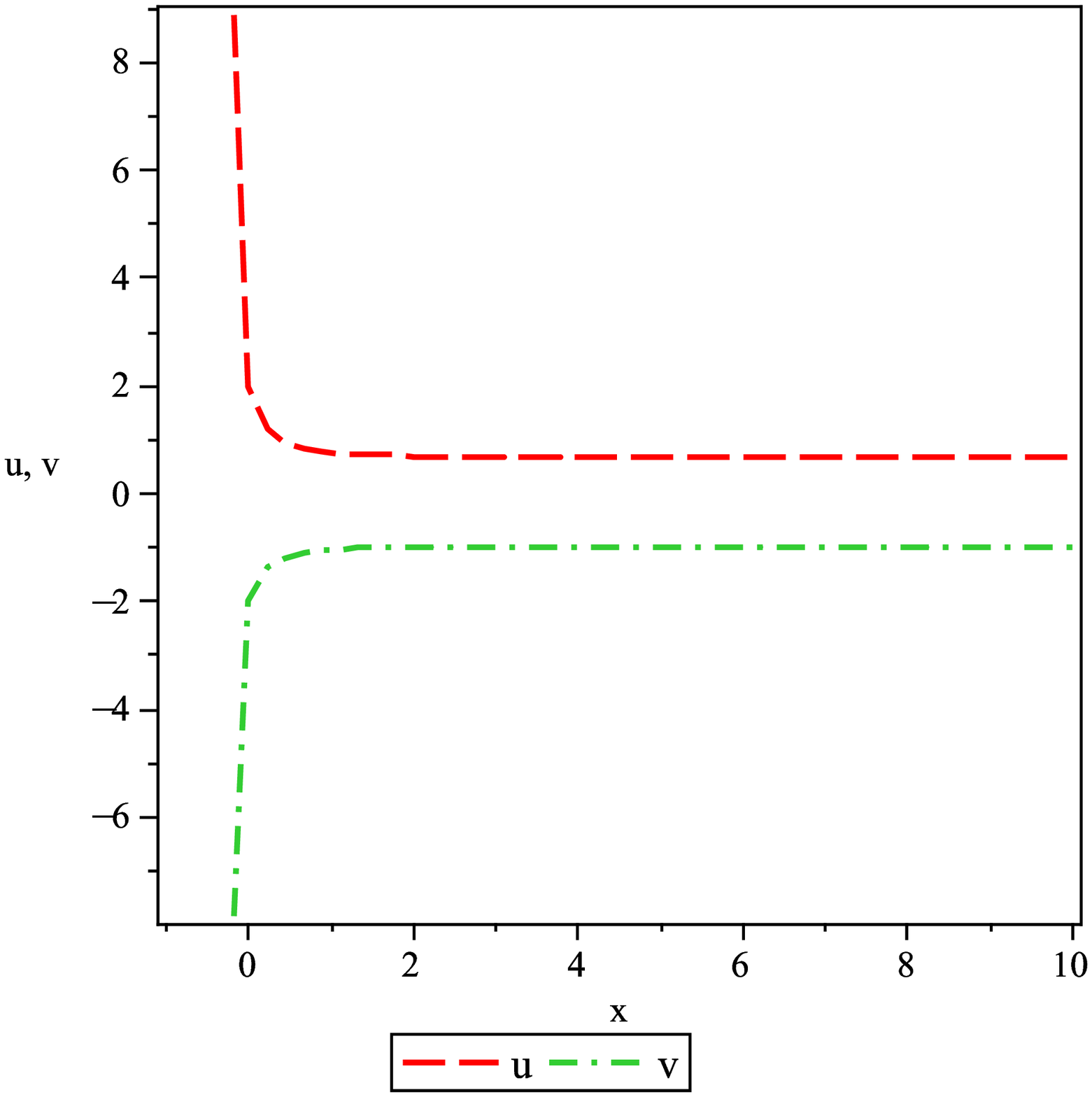}\\
\caption{The parametric functions $u$ and $v$ are plotted against
the e-folding time parameter $x$. The parameters are fixed as
$c=0.3$, $\alpha=0.4$ and $\omega_X=-1$. The initial condition is
$u(0)=2, v(0)=-2$. }
\end{figure}
\newpage
\begin{figure}
\includegraphics[scale=.7]{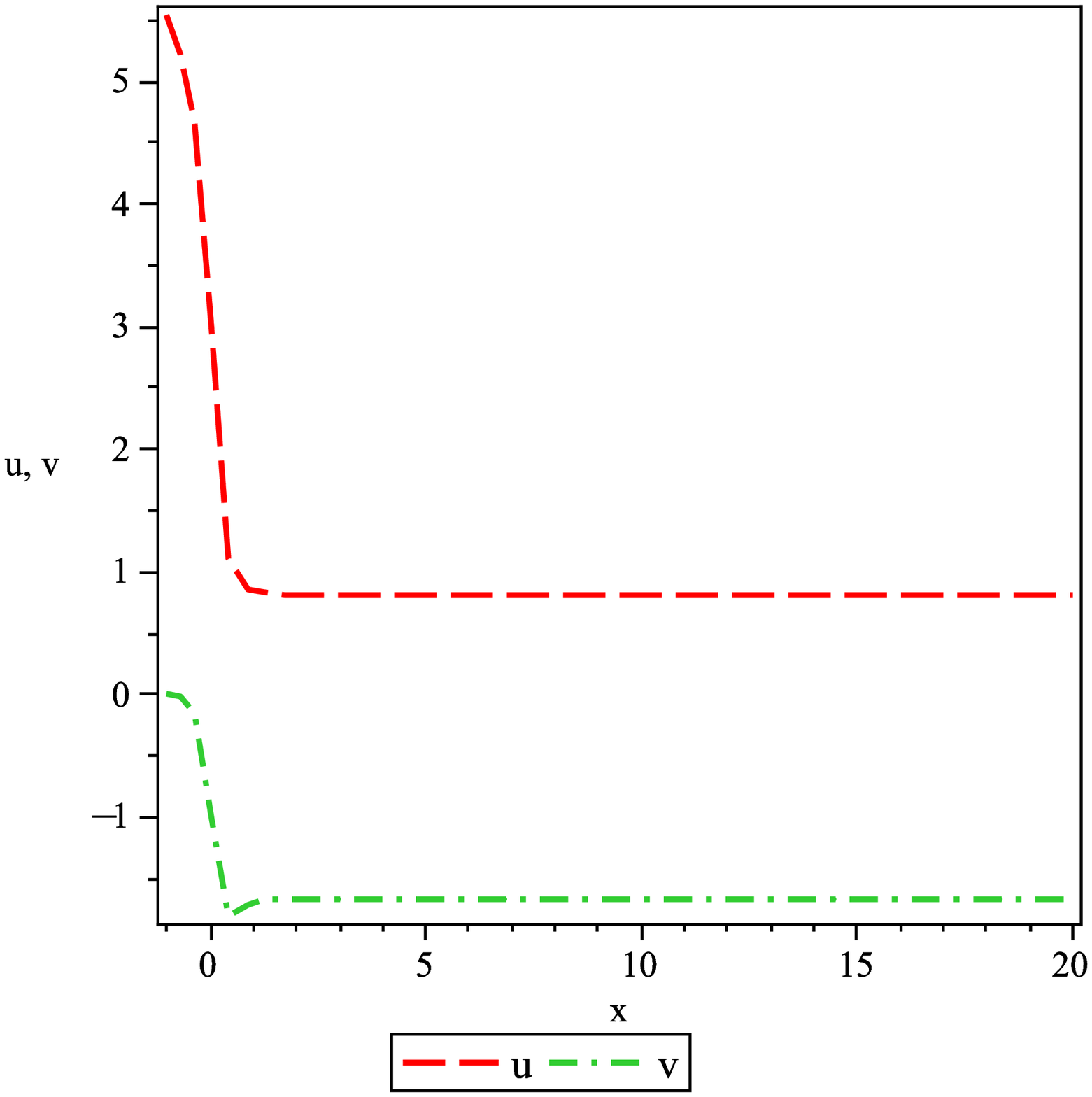}\\
\caption{The parametric functions $u$ and $v$ are plotted against
the e-folding time parameter $x$. The parameters are fixed as
$c=0.5$, $\alpha=0.08$ and $\omega_X=-1/3$. The initial condition is
$u(0)=3, v(0)=-1$.}
\end{figure}
\newpage
\begin{figure}
\includegraphics[scale=.7]{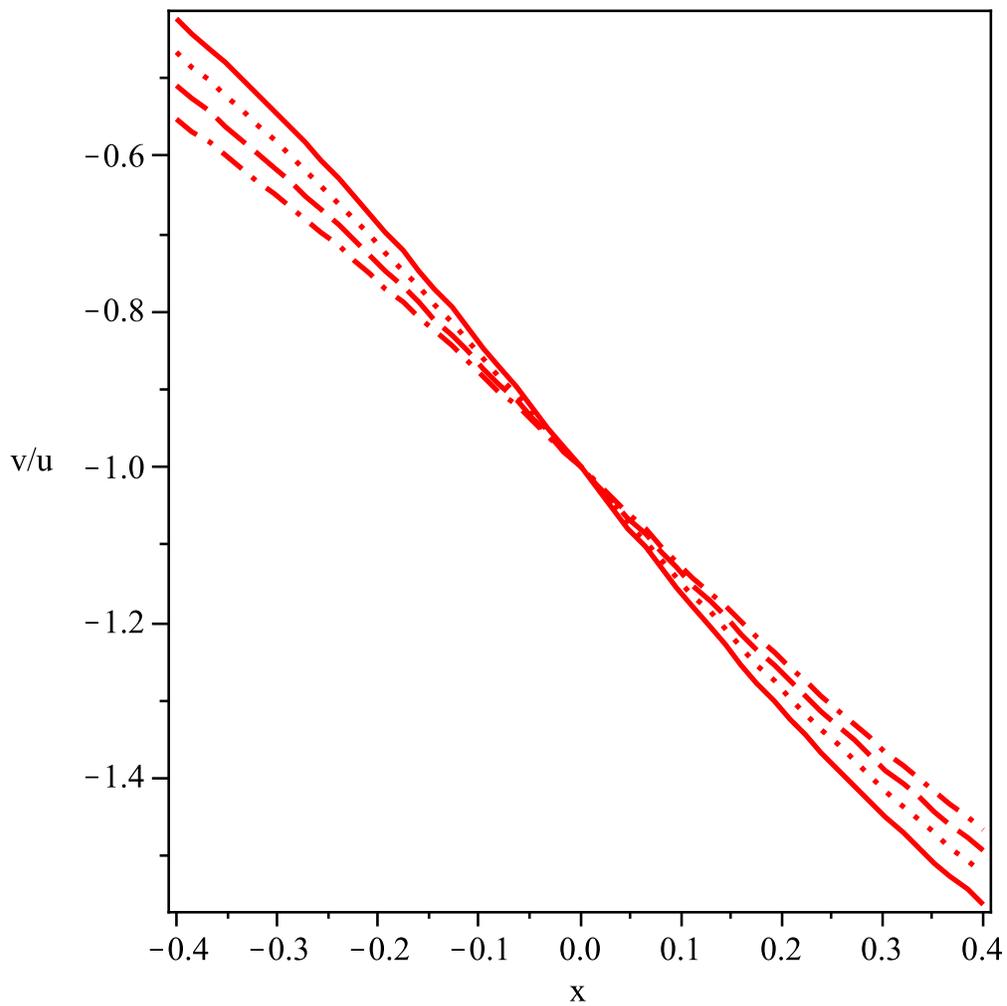}\\
\caption{The EoS parameter $\omega_X=v/u$ is plotted against $x$.
The parameters are fixed as $c=0.5$, $\alpha=0.08$ and
$\omega_X=-1$. The initial conditions are $u(0)=1, v(0)=-1$ (solid
line); $u(0)=1.1,v(0)=-1.1$ (dots); $u(0)=1.2,v(0)=-1.2$ (dashes);
$u(0)=1.3,v(0)=-1.3$ (dash dot).}
\end{figure}
\newpage
\begin{figure}
\includegraphics[scale=.7]{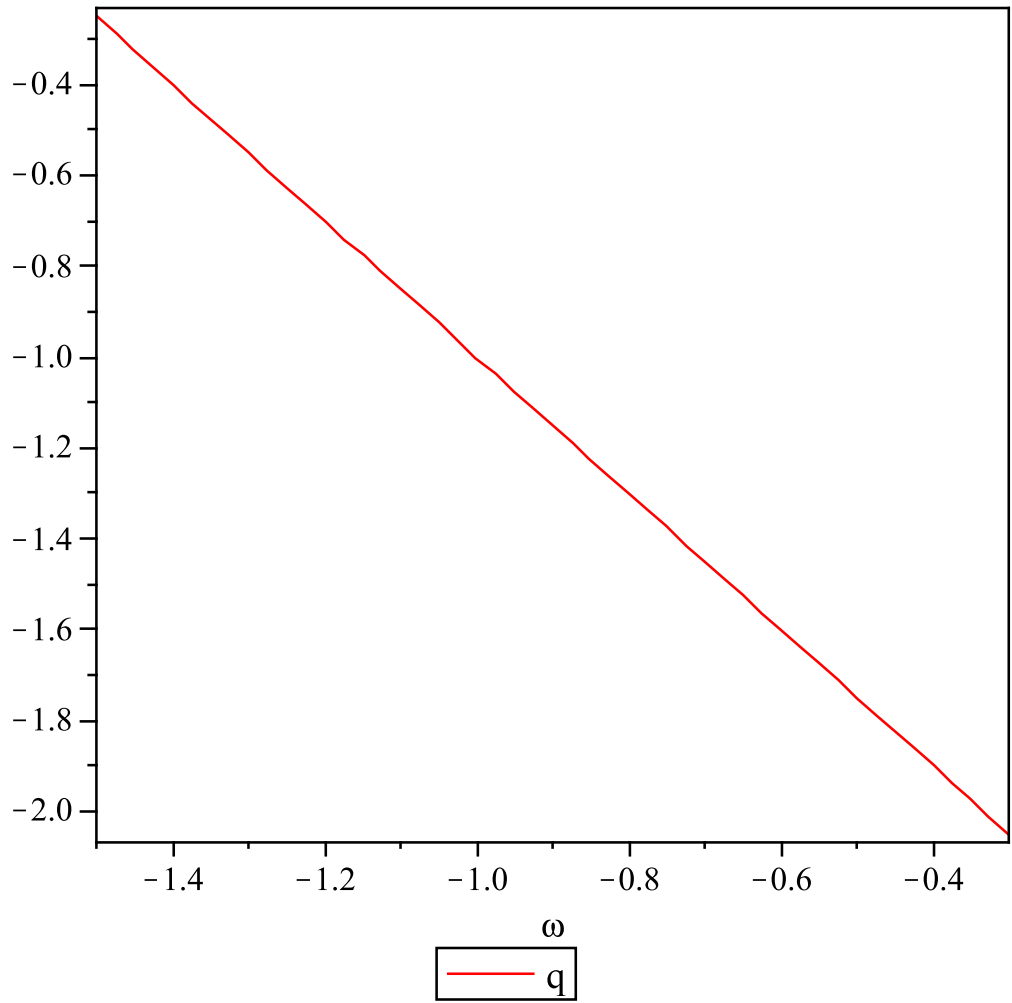}\\
\caption{The deceleration parameter $q$ is plotted against $x$.}
\end{figure}
\end{document}